\documentclass[twocolumn,showkeys,preprintnumbers,amsmath,amssymb]{revtex4}

\usepackage{graphicx}
\usepackage{dcolumn}
\usepackage{bm}
\usepackage[latin1]{inputenc}

\begin{document}


\title{
Modeling of a Resonant Tunneling Diode Optical Modulator}

\author{J. J. N. Calado and J. M. L. Figueiredo}
 \email{jlongras@ualg.pt}
\affiliation{Departamento de Física, Faculdade de Ciências e Tecnologia, Universidade do Algarve,
Campus de Gambelas, 8005-139 Faro, Portugal
}%
\author{C. N. Ironside}
\affiliation{ Department of Electronics and electrical Engineering, University of Glasgow G12 8LT,
UK
}%


\begin{abstract}
The integration of a double barrier resonant tunneling diode within a unipolar optical waveguide
provides electrical gain over a wide bandwidth. Due to the non-linearities introduced by the double
barrier resonant tunneling diode an unipolar InGaAlAs/InP optical waveguide can be employed both as
optical modulator and optical detector. The modeling results of a device operating as optical
modulator agree with preliminary experimental data, foreseeing for an optimized device modulation
depths up to 23 dB with chirp parameter between -1 and 0 in the wavelength range analyzed (1520 nm
- 1600 nm).
\end{abstract}


\maketitle

\section{Introduction}
Due to their high-speed response and radio frequency (rf) gain, several groups have proposed the
application of resonant tunnelling diodes (RTDs) in the optical and infrared domains \cite{Mizuta}.

A novel optoelectronic device based on the integration of a RTD with an optical waveguide (OW), the
RTD-OW, as been proposed for electro-optical and opto-electrical conversion \cite{McMeekin}. The
waveguide configuration is used to ensure large interaction volume between the RTD regions and the
guided light. The full demonstration and development of this new device concept in the ternary
AlGaAs/GaAs and quaternary InGaAlAs/InP material systems can be of great importance for optical
communication, specially in high-speed fibre radio links. The AlGaAs/GaAs material system is
interesting for short hall communications in the wavelength range around 900 nm, and the
InGaAlAs/InP material system is useful in the wavelength range where optical fibres have the lowest
loss and chromatic dispersion (1300 nm to 1600 nm).

Our group has successfully integrated AlGaAs and InGaAlAs RTDs within unipolar AlGaAs/GaAs and
InGaAlAs/InP optical waveguides, respectively, and demonstrated optical modulation of guided light
around 900 nm and 1550 nm \cite{Figueiredo1}\cite{Figueiredo2}. In this paper we report modeling
results of the device operating as an electro-optical converter at 1550 nm, confirms preliminary
experimental data \cite{Figueiredo2}\cite{Figueiredo3}.

\section{Principle of operation as a modulator}

\begin{figure}[hbt]
\includegraphics[width=0.475\textwidth]{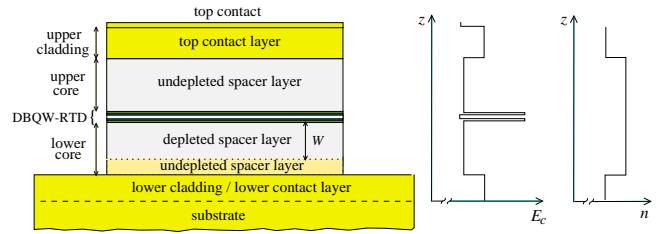} \caption{\label{RTD-OPT_WG_1}Schematic diagram of the
RTD-OW wafer structure, \protect\( \Gamma \protect \)-conduction band-edge and refractive index
profiles.}
\end{figure}

Essentially, the RTD-OW is a unipolar device consisting of a double barrier quantum well (DBQW)
resonant tunneling diode (RTD) embedded within a ridge channel optical waveguide. The DBQW-RTD
structure consists of two AlAs barriers surrounding a InGaAs quantum well. The optical waveguide
comprehends two moderately n-doped InGaAlAs layers lattice matched to InP, sandwiched between
highly n-doped InP cladding regions, Fig. \ref{RTD-OPT_WG_1}.

\begin{figure}[hbt]
\includegraphics[width=0.425\textwidth]{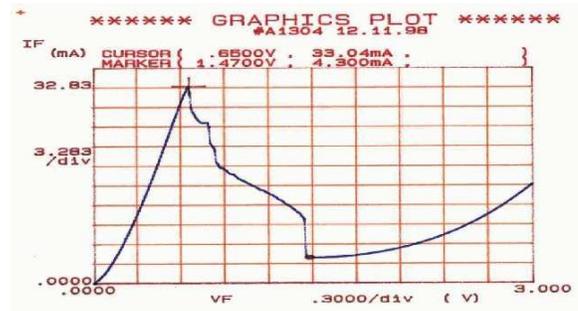}\caption{\label{PCVR-2} Experimental current-voltage
characteristic of an InGaAlAs/InP RTD-OW.}
\end{figure}

The presence of the DBQW-RTD within the waveguide core introduces high nonlinearities in the
current-voltage (I-V) characteristic of the unipolar waveguide: the device I-V curve shows large
negative differential resistance (NDR), Fig. \ref{PCVR-2}. The physics that gives rise to this type
of I-V is explained in \cite{Figueiredo2}.

\begin{figure}[hbt]
\includegraphics[width=0.475\textwidth]{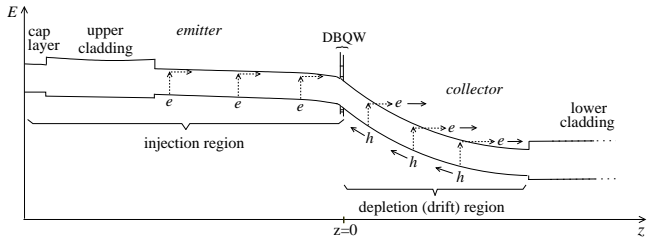} \caption{\label{RTD-OW-FKEF_1}Schematic diagram
of the energy bands in a RTD-OW at the valley voltage as a function of distance: upper curve is the
lowest conduction band energy and the lower curve represents the highest valence band energy. The
light absorption occurs mainly in the device depletion region.}
\end{figure}

As a consequence of the presence of the RTD within the unipolar waveguide core a non-uniform
potential distribution is induced across the waveguide cross-section, Fig. \ref{RTD-OW-FKEF_1}. The
magnitude of the associated electric field distribution depends strongly on the bias voltage. When
the device operating point switches from the peak to the valley regions of the I-V curve there is
an enhancement of the electric field across the waveguide depletion region. This produces
substantial changes in the absorption coefficient of the waveguide at wavelengths near the core
material band-edge via the Franz-Keldysh effect (for more details see \cite{Figueiredo2}). The
operation of the RTD-OW as an optical modulator takes advantage of this absorption change induced
by the RTD peak-to-valley switching.

\begin{figure}[hbt]
\includegraphics[width=0.425\textwidth]{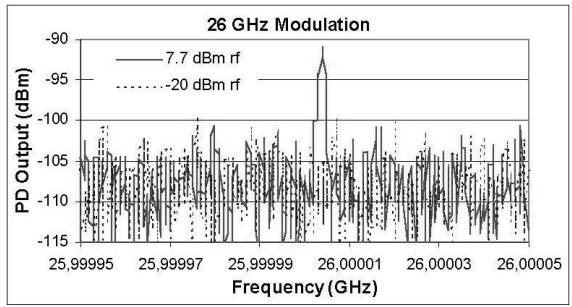} \caption{\label{26GHz} Electrical spectra of optical signals
modulated at 26 GHz for driving powers of -20 dBm and +7.7 dBm.}
\end{figure}

Preliminary experimental results indicate that a non-optimized device designed to act as a
modulator is capable of high-speed modulation (up to 26 GHz) with extinction ratio higher than 10
dB over a wide range of wavelengths at driving power as low as 7.7 dBm, Fig. \ref{26GHz}
\cite{Figueiredo3}.

\section{Device modeling phases}

The device modeling comprehends two phases. In the first, the electrical characteristics of the
device such as the I-V curve and the potential distribution across the waveguide cross-section are
determined. In the second part, the waveguide optical properties induced changes such as the
extinction coefficient and the refractive index variations at different bias voltage and as
function of the guided light wavelength are calculated.

\begin{figure}[hbt]
    \center
\includegraphics[width=0.375\textwidth]{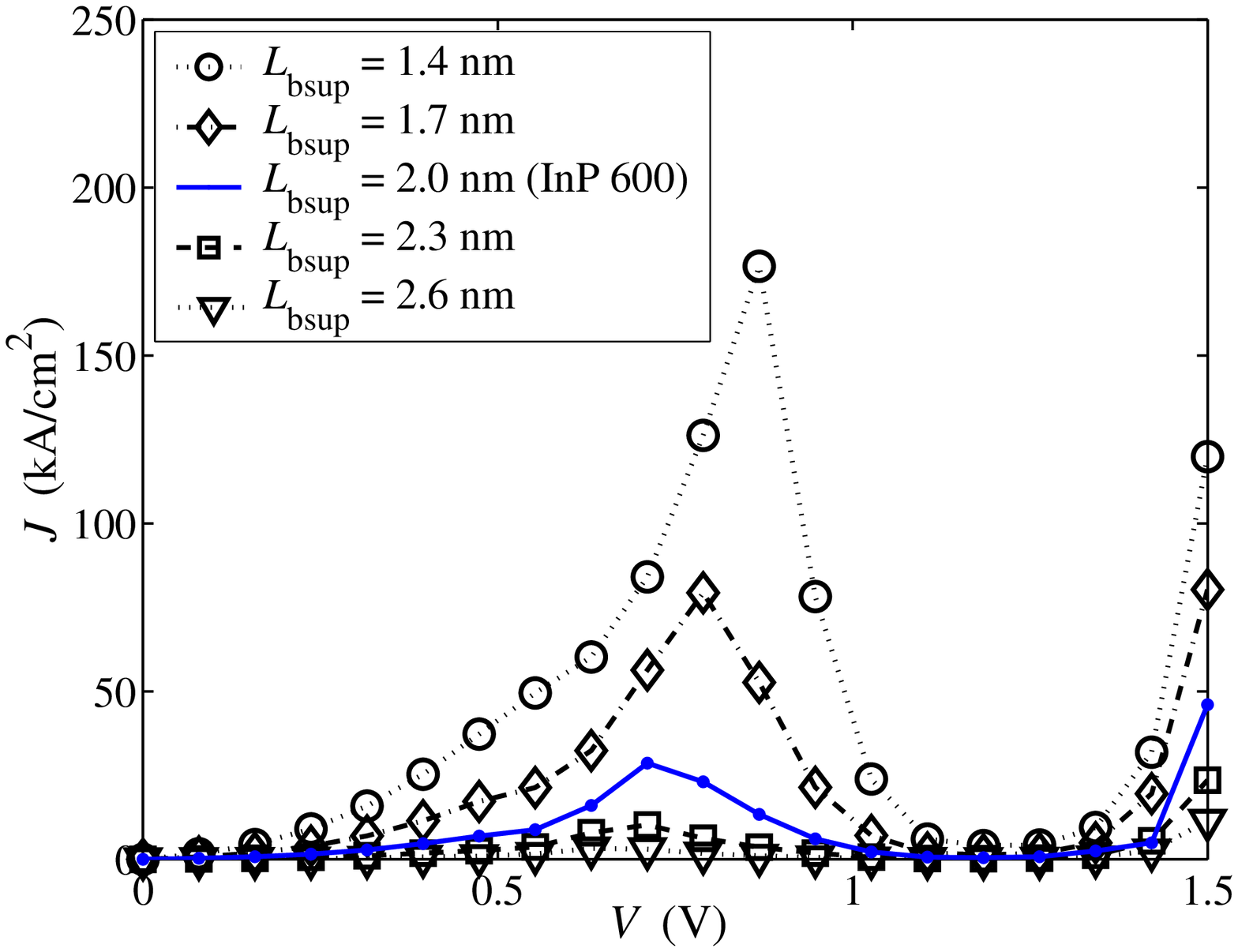}\\(a)\\
\includegraphics[width=0.375\textwidth]{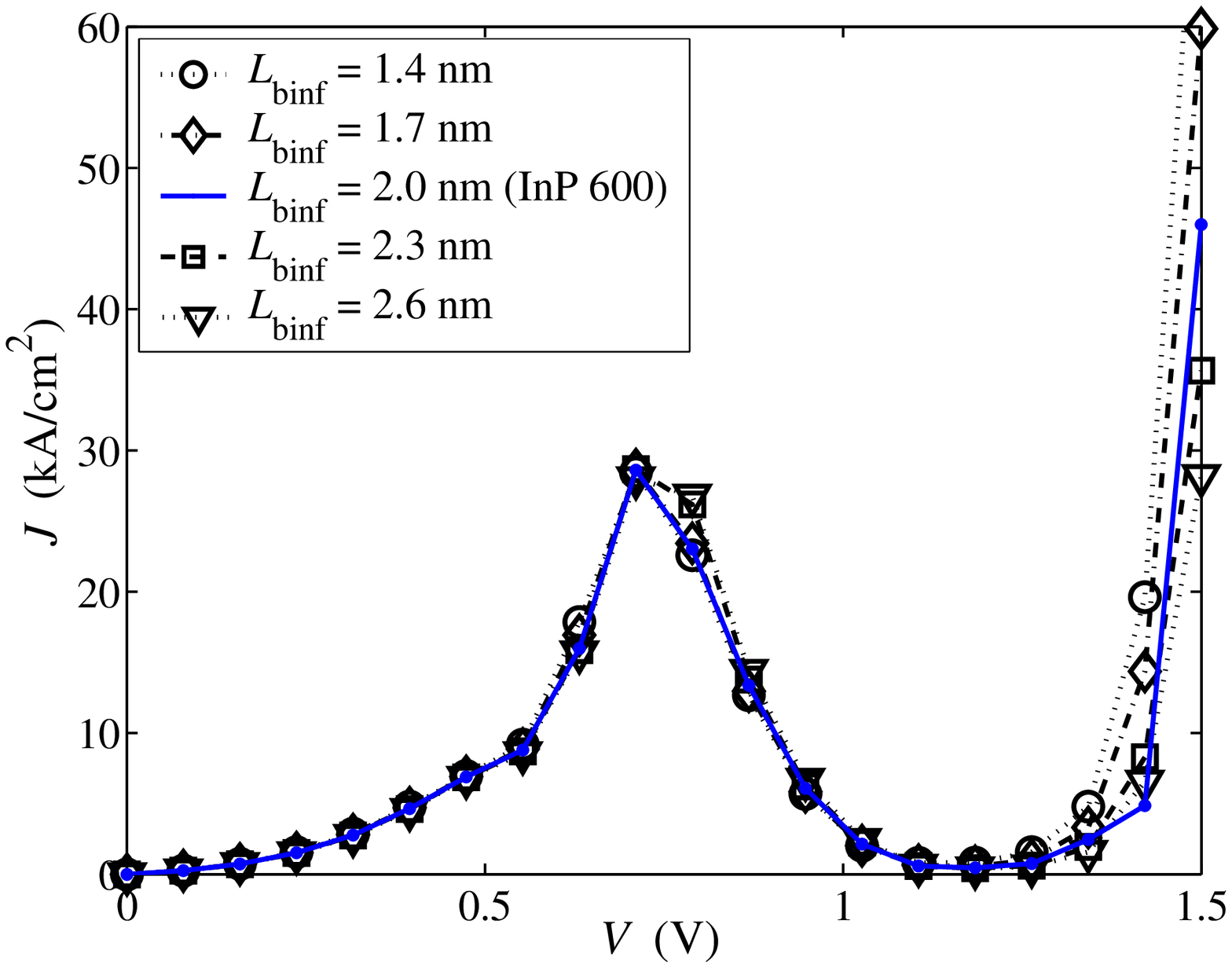}\\(b)\caption{\label{Dopagem-JV} Typical device I-V characteristics
determined by the WinGreen simulation package for several barrier thickness: a) emitter barrier and
b) collector barrier (see Fig. \ref{RTD-OW-FKEF_1}).}
\end{figure}

The RTD-OW electrical modelling employs the WinGreen simulation package \cite{WinGreen} that
determines the device I-V characteristic and the potential profiles across the waveguide
cross-section as function of the applied bias. The I-V curve permits to extract the NDR region
characteristics: the peak and the valley voltages, Fig. \ref{Dopagem-JV}.

\begin{figure}[hbt]
    \center
\includegraphics[width=0.235\textwidth]{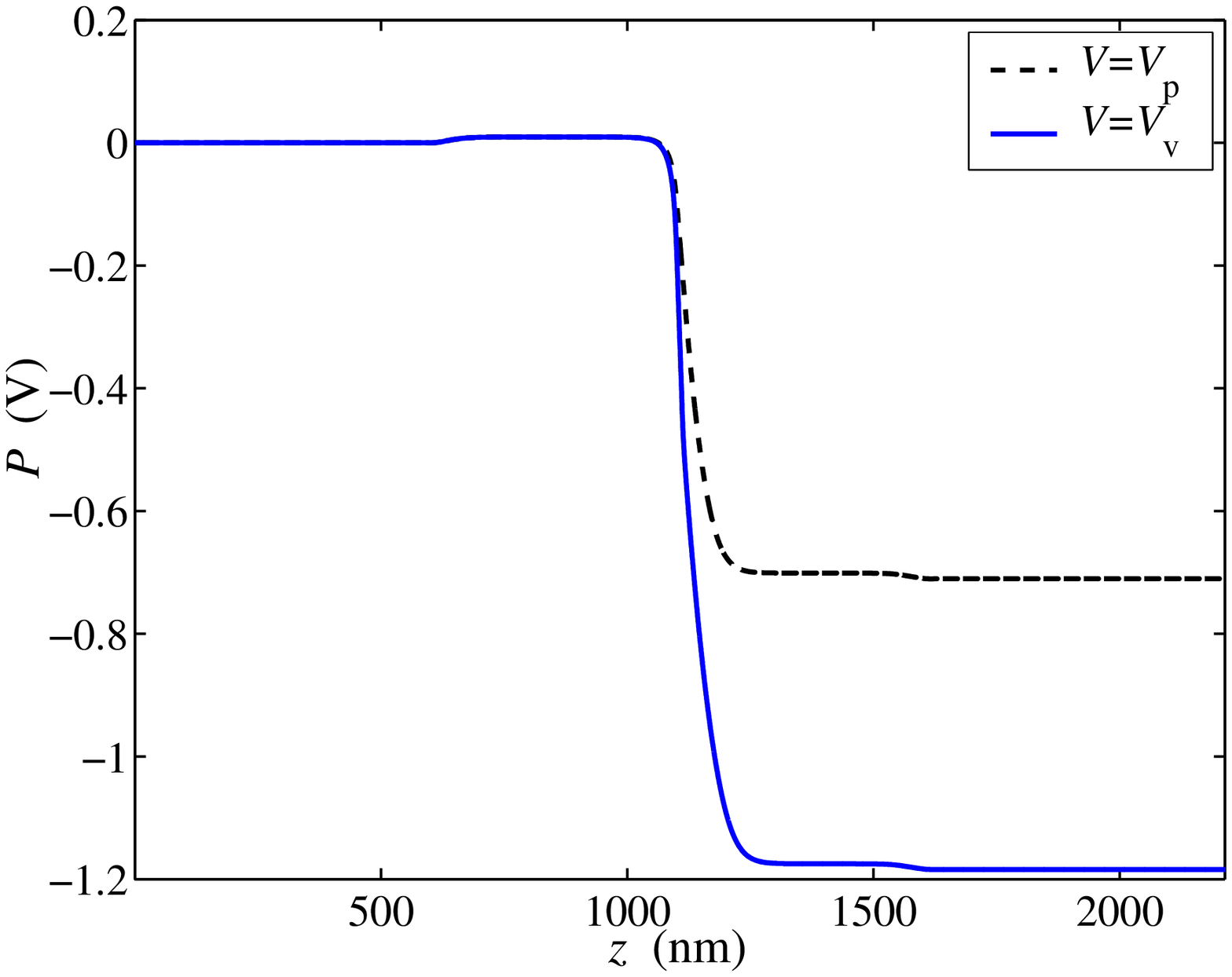}
\includegraphics[width=0.235\textwidth]{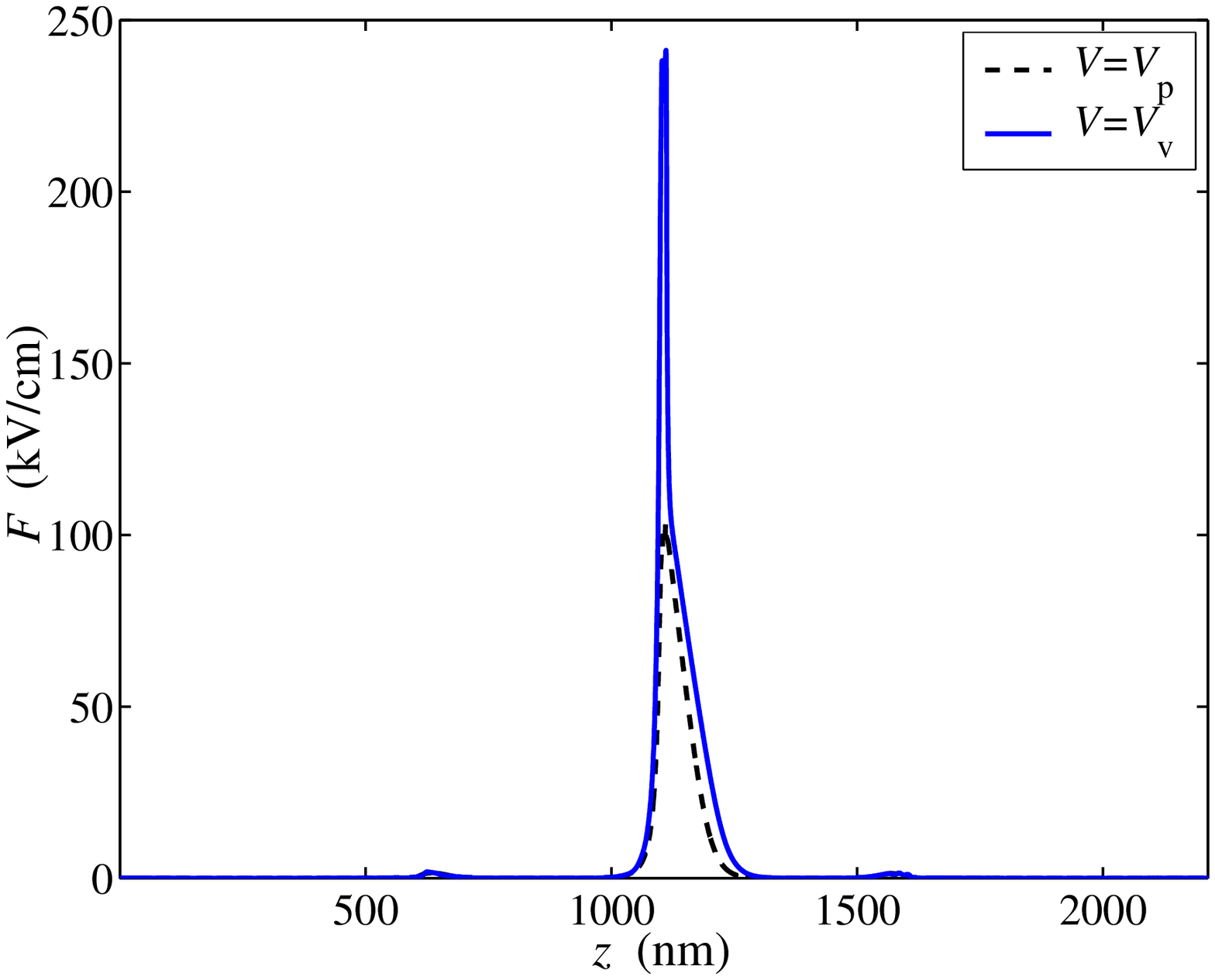}\\(a) ~ ~  ~ ~ ~ ~ ~ ~ ~ ~ ~ ~ ~ ~ ~ ~ ~(b)\caption{\label{Campo-Electrico}
The device typical potential distributions at the peak and the valley voltages a) and the
corresponding electric field distributions b).}
\end{figure}

The electric field distribution \(F(z)\) across the waveguide for a given bias voltage \(V\), Fig.
\ref{Campo-Electrico}b), is obtained through the gradient of the potential distribution, Fig.
\ref{Campo-Electrico}a).

The waveguide absorption coefficient at a given voltage \(V\) is then determined as a function of
the electric field distribution \(F(z)\), in V/cm, and of the light energy \(E\), in eV, using the
Franz-Keldysh relation \cite{Chuang},
\begin{equation}\label{alpha}
\alpha(E,
F)=\sum_{j}A_{j}F^{\frac{1}{3}}\left[\left|\left(\frac{d\textrm{Ai}}{dz}\right)_{\beta_{j}}\right|^{2}-
\beta_{j}\left|\textrm{Ai}(\beta_{j})\right|^{2}\right],
\end{equation}
where Ai is the Airy function, \(A_{j}=7,65\times10^{5}C(2m_{rj}/m_{0})^{\frac{4}{3}}/nE\),
\(\beta_{j}=B_{j}(E_{g}-E)F^{-\frac{2}{3}}\),
\(B_{j}=1,1\times10^{5}(2m_{rj}/m_{0})^{\frac{1}{3}}\), \(E_{g}\) is the core band-gap energy
expressed in eV, and \(n\)  is the refractive index; \(m_{rj}\)  and \(m_{0}\) are the
electron-hole reduced effective mass and electron rest mass, respectively; \(C\) is a scaling
parameter to adjust Eq. \ref{alpha} to independent experimental data \cite{Chuang}. The sum is over
the light and heavy holes.

The absorption change due to the RTD peak to valley switching is then given by:
\begin{equation}\label{delta-alpha}
\Delta\alpha(\hbar\omega, \Delta F_{p-v})=\alpha(\hbar\omega, F_{v})-\alpha(\hbar\omega, F_{p}),
\end{equation}
where \(F_{p,v}\) represents the magnitude of the electric field at the peak (p) and at the valley
(v), respectively.

The refractive index change as function of the bias is determined from the absorption change
through the Kramers-Kronig relation \cite{Chuang}
\begin{equation}\label{Kramers-Kronig}
\Delta n(\hbar\omega, F)=\frac{\hbar c}{\pi}\int_{0}^{\infty}\frac{\Delta\alpha(\hbar\omega',
 F)}{(\hbar\omega')^{2}-(\hbar\omega)^{2}}d\omega',
\end{equation}
The modulation depth is estimated through
\begin{equation}\label{modulation-depth}
R_{on-off}(\textrm{dB})=4,343\gamma_{f}\Delta\alpha,
\end{equation}
where \(\gamma_{f}\) represents the overlap integral between the electric field and the optical
field  distributions. The chirp parameter is calculated using the relation
\begin{equation}\label{chirp-parameter}
\alpha_{H}=\Delta n/\Delta k,
\end{equation}
where \(k\) is the extinction coefficient, given by \(\Delta k=\Delta\alpha \lambda/4\pi \).

\section{Modeling results}

The objective of this work is to understand the role of the structural and material parameters on
the device performance in order to determine the optimized device structure for each application.
The study reported here aims the determination of the RTD-OW structure that gives the highest
modulation depth with the minimum propagation loss. To achieve this purpose it is necessary to
maximize the electric field change when the device operation point switches from the peak to the
valley region and the overlap between the electric field distribution and the guided mode.

\begin{figure}[hbt]
    \begin{center}
    \includegraphics[width=0.375\textwidth]{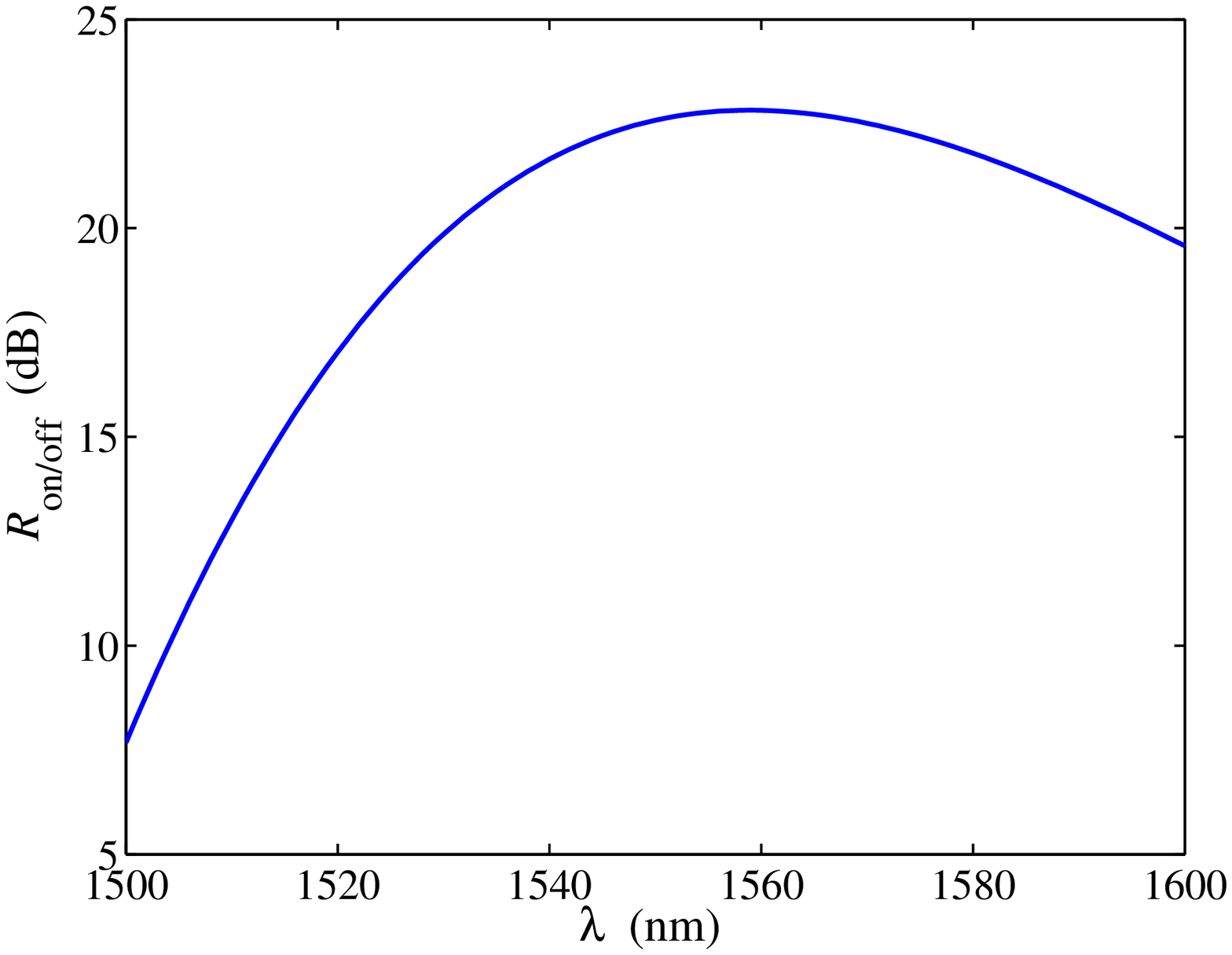}\\(a)\\
    \includegraphics[width=0.375\textwidth]{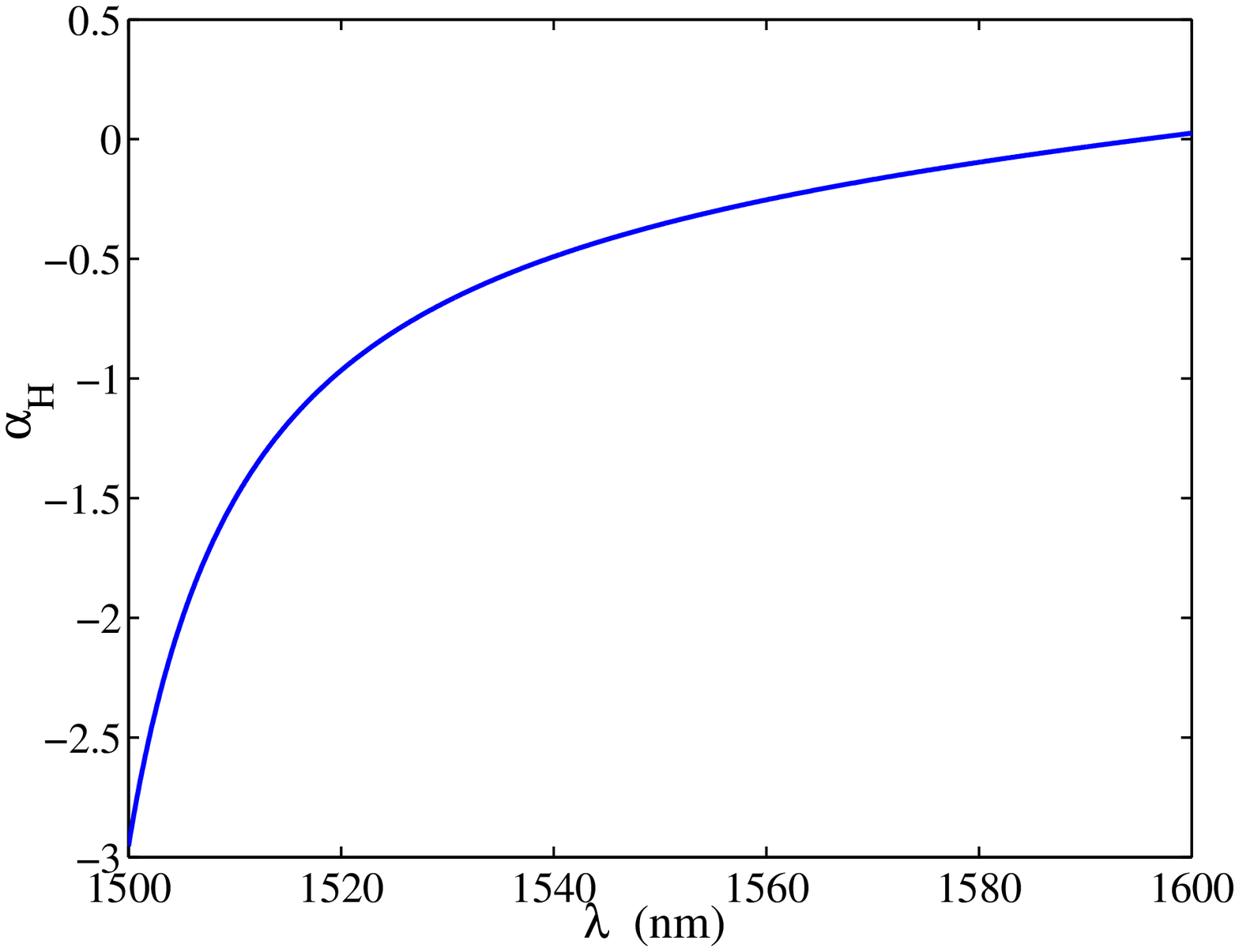}\\(b)
    \caption[Razão de extinção, parâmetro de \emph{chirp},
    alteração de índice de refracção e variação do coeficiente de
    extinção, para a estrutura RTD-EAM optimizada.]
    {(a) Extinction ratio $R\rm_{on/off}$ and (b) chirp parameter $\alpha\rm_H$
    as function of the wavelength induced by the peak-to-valley switching.}
    \label{R-Chirp-optim}
    \end{center}
\end{figure}

According the modeling results, the optimized modulator structure corresponds to a RTD-OW
configuration consisting of two 2 nm thick AlAs barriers surrounding a 6 nm thick InGaAlAs quantum
well, symmetrically embedded within a 300 mm unipolar InGaAlAs waveguide core with doping
concentration of \(1\times10^{16}\) cm\(^{-3}\) and 600 nm thick InP cladding layers highly doped
(\(2\times10^{18}\) cm\(^{-3}\)). A 200 \(\mu\)m long RTD-OW shows extinction ratio up to 23 dB
around 1560 nm with chirp parameter of -0.25, Fig. \ref{R-Chirp-optim}. The results also show the
propagation loss are considerable higher than the ``best" values reported in the literature. The
way to decrease the propagation loss are under current investigation.

\begin{figure}[hbt]
    \begin{center}
    \includegraphics[width=0.375\textwidth]{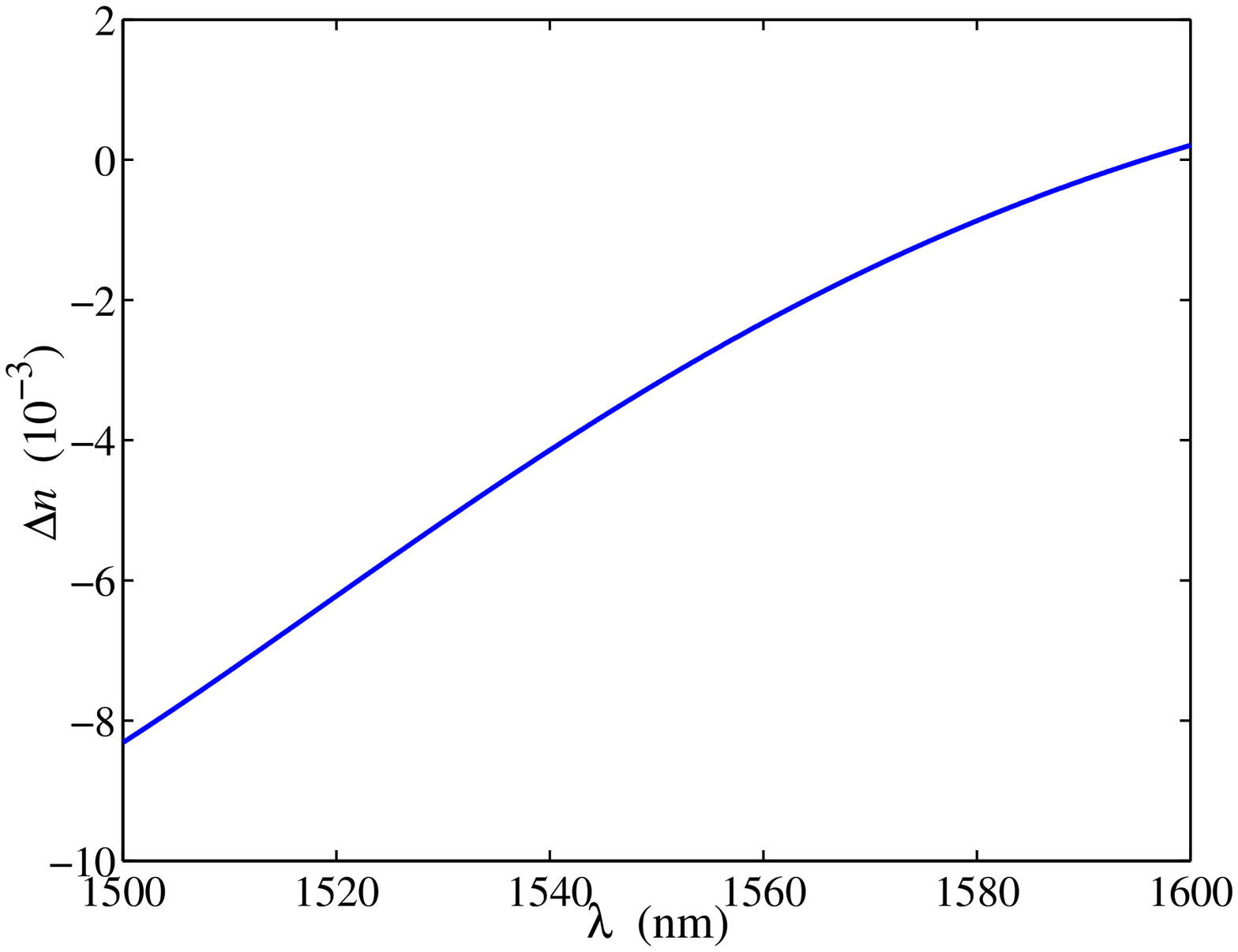}\\(a)\\
    \includegraphics[width=0.375\textwidth]{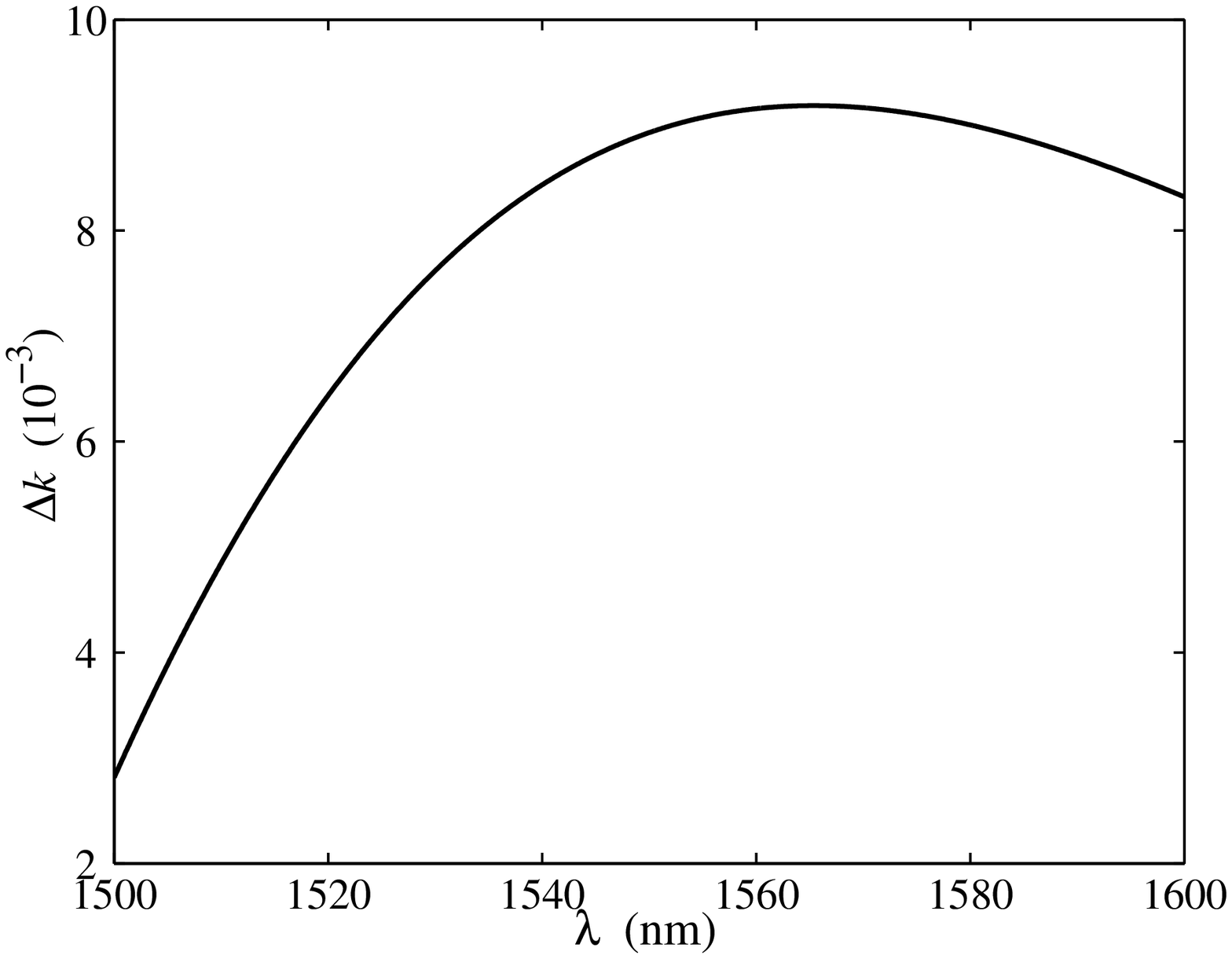}\\(b)\\
    \caption[Razão de extinção, parâmetro de \emph{chirp},
    alteração de índice de refracção e variação do coeficiente de
    extinção, para a estrutura RTD-EAM optimizada.]
    {(a) Refractive index variation $\Delta n$ and (b) extinction coefficient change $\Delta k$
    as function of the wavelength induced by the peak-to-valley switching.}
    \label{N-K-optim}
    \end{center}
\end{figure}

The refractive index change and the extinction coefficient variation are shown in Fig.
\ref{N-K-optim}, and appears to indicate the RTD-OW has high potential to operate as an
electro-refraction modulator through an interferometer configuration.



\section{Conclusion and future work}

The modeling results corroborate preliminary modulation depth experimental data
\cite{Figueiredo2}\cite{Figueiredo3}. The fabrication of devices based on the present optimized
structures is being considered.

The current devices also show detection capabilities. Its is foreseen the device can operate as a
light detector incorporating an intrinsic integrated amplifier for the photocurrent generated by
the incident light. The presented model is currently being extended to include the effect of light
absorption.

\section*{Acknowledgment}
This work is supported by the Fundação para a Ciência e a Tecnologia (Portugal) under Grant No.
POSI/41760/CPS/2001.

\end{document}